\documentclass[a4paper,amsfonts,amssymb,amsmath,reprint,showkeys,nofootinbib,twoside]{revtex4-1}

\usepackage[utf8]{inputenc}
\usepackage[english]{babel}
\usepackage{amsmath}
\usepackage{graphicx}
\usepackage{mathtools}

\usepackage[colorinlistoftodos, color=green!40, prependcaption]{todonotes}

\usepackage{amsthm}
\usepackage{mathtools}
\usepackage{physics}
\usepackage{xcolor}
\usepackage{graphicx}
\usepackage[left=23mm,right=13mm,top=35mm,columnsep=15pt]{geometry} 
\usepackage{adjustbox}
\usepackage{placeins}
\usepackage[T1]{fontenc}
\usepackage{lipsum}
\usepackage{csquotes}

\usepackage[colorlinks=true]{hyperref}

\begin{document}

\title{Robustness of bound states in the continuum in metasurface based on Ge$_2$Sb$_2$Te$_5$ versus structural imperfections}

\author{Nikolai A. Vlasov$^{1}$, Alexander I. Solomonov$^{1}$, Zarina F. Kondratenko$^{1}$, Kirill A. Bronnikov$^{1}$, Mikhail V. Rybin$^{1, 2}$, and Ekaterina E. Maslova$^{1}$}
\email[Correspondence email address: \\]{ekaterina.maslova@metalab.ifmo.ru}
\affiliation{$^1$School of Physics and Engineering, ITMO University, 197101, St. Petersburg, Russia\\
$^2$Ioffe Institute, 194021, St. Petersburg, Russia\\
}

\date{\today}

\begin{abstract}
We study the impact of lithography imperfections on quasi-bound states in the continuum (quasi-BICs) supported by a one-dimensional metasurface of Ge$_2$Sb$_2$Te$_5$ (GST) bars with trapezoidal deviations from rectangular cross-sections. Several mechanisms of quality ($Q$) factor scaling, including the impact of material losses, dispersion, and geometric imperfections are established. We demonstrate that transition to identical isosceles trapezoids, despite preserving the required $C_2$ symmetry, reduces the $Q$ factor in the amorphous phase due to absorption changes accompanying the resonance shift. 
Further, the $Q$ factor remains robust for both GST phases under random element-to-element variations of the trapezoid angle, while analytical and numerical estimations in the absence of material losses show inverse-quadratic scaling of the Q factor with the disorder amplitude. We reveal that in the GST-based metasurface, the $Q$ factor is tolerant to geometric imperfections for insignificant dispersion near the BIC wavelength, but changes in case of substantial dispersion. The phase shifting and established robustness of BICs in GST can be useful for applications where stable moderate $Q$ factors are essential.

\end{abstract}


\maketitle

\section{Introduction}
\label{Introduction}

Bound states in the continuum (BICs) are spatially localized states whose energies lie within the continuous spectrum, yet they remain decoupled from continuum radiation channels. In photonics, BICs can be realized in periodic high-index structures with a refractive-index modulation~\cite{Hsu2016Jul, Koshelev2020Engineering}. In an ideal case, BICs have zero radiative loss, and, consequently, their $Q$ factor becomes infinite. However, in practice, material absorption, finite-size effects, and fabrication-induced geometric imperfections~\cite{maslova2021bound, sadrieva2018experimental, bulgakov2019high, Sidorenko2021Mar, Semushev2026RobustnessBICBilayer} turn ideal BICs into quasi-BICs (q-BICs) with a finite $Q$ factor.

Among dielectric platforms for BIC photonics, phase-change materials (PCMs) are particularly promising because their optical properties can be reversibly reconfigured between two stable states. Ge$_2$Sb$_2$Te$_5$ (GST) is a prototypical PCM, with a complex refractive index changing upon a transition between the amorphous and crystalline phases, creating a pronounced contrast in both refraction and absorption~\cite{Singh2013May, Frantz2023Dec, Vincenti2024Aug}. This large modulation of the dielectric function directly affects the modal structure of a GST-based metasurface, where the resonance wavelength, field confinement, and $Q$ factor of a supported mode are all defined by the permittivity of the constituent material. As a result, the same lithographic geometry can sustain a BIC-like resonance in both phases, with spectral position and $Q$ factor determined solely by the GST phase without any geometric reconfiguration. This property makes GST an ideal candidate for BIC tuning and highly relevant to reconfigurable photonic devices {such as tunable filters, sensors, and optical switches, where the ability to shift or suppress a BIC on demand is essential~\cite{Singh}.}

Realizing symmetry-protected BICs requires lithographic fabrication of volumetric dielectric or semiconductor metasurfaces~\cite{Zhao2023, Park2023UV}, which inherently introduces unavoidable deviations from the nominal specifications. Moreover, the etching process creates resonant elements with non-uniform geometry~\cite{Solomonov2023, Park2023UV, Roadmap2024}. These deviations are especially critical for symmetry-protected BICs, whose nature relies on the $C_2$ symmetry of the unit cell~\cite{Koshelev2018Nov, maslova2021bound}. Any random element-to-element variation of the unit cell geometry breaks this symmetry and transforms an ideal BIC into a finite-$Q$ resonance~\cite{Koshelev2018Nov, Ni2017Mar, maksimov2025temporal}. The problem is further complicated in GST-based structures, where the strong dispersion and absorption add an intrinsic material loss channel that competes with disorder-induced losses.

\begin{figure*}[th]
\centering
\includegraphics[width = 2\columnwidth]{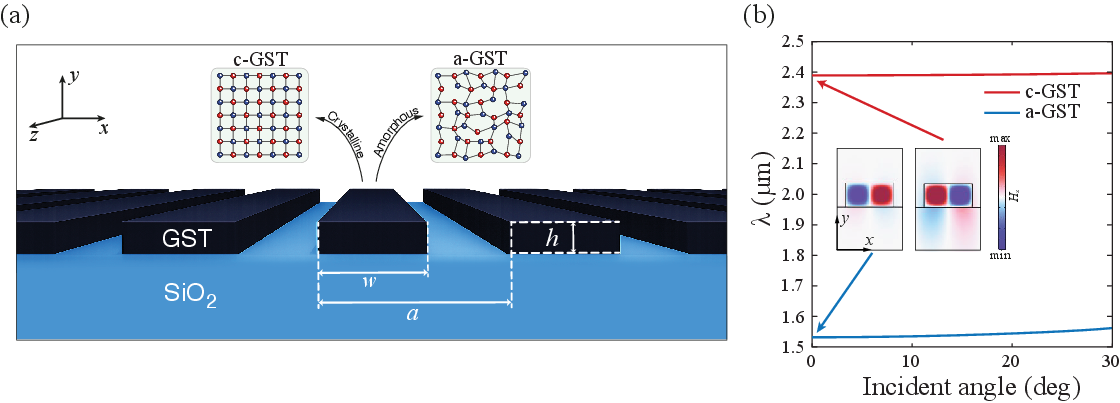}
\caption{(a) Schematic illustration of the GST-based structure. Black bars represent GST elements, while the glass substrate is shown in blue. The period of the structure is $a$, and $w$ and $h$ are the width and height of GST bars, respectively.
(b) Dispersion of q-BICs supported by crystalline- and amorphous-phase GST bars for angles of incidence from $0^\circ$ to $30^\circ$. Blue line corresponds to the crystalline phase of GST, and red line to the amorphous phase. Insets show distributions of the $H_z$ component amplitude of the magnetic field. }
\label{Intro}
\end{figure*}

In this work, we present a theoretical study of symmetry-protected BICs supported by an array of GST bars in the crystalline (c-GST) and amorphous (a-GST) phases.
We focus on the effects of trapezoidal cross-sections, which act as deviations from the ideal rectangular ones.
Such trapezoidal cross-sections are nearly unavoidable due to lithography etching process and can significantly affect the $Q$ factor of BICs. {Therefore, the $Q$ factor in GST-based metasurfaces is affected by complex interplay between material losses, frequency dispersion and geometric imperfections.}

{Our results show that for c-GST, isosceles trapezoidal deformations have only a minor effect on the Q factor, whereas in a-GST, they reduce Q due to strong dispersion near the BIC wavelength. By analyzing fluctuations in the geometric parameters, we find both analytically and numerically that, in the absence of material loss, the Q factor follows an inverse-quadratic dependence on the disorder amplitude and on specific geometric parameters. More broadly, this inverse-quadratic scaling is material-independent and applies to high-index dielectric metasurfaces in which sidewall-angle fluctuations are the dominant fabrication imperfection. Once intrinsic material loss is accounted for, it becomes the dominant factor over geometric disorder, underscoring the robustness of the $Q$ factor in GST-based nanostructures. This combination of disorder resilience and phase-switching capability makes GST promising for moderate-$Q$ BIC applications such as tunable filters, optical switches, modulators, and reconfigurable sensing platforms, where stable resonant response and on-demand spectral tuning are more important than achieving the highest possible $Q$ factor.}

\section{Structure}

We consider a periodic array of infinitely long GST bars with a rectangular cross-section on a SiO$_2$ substrate [Fig.~\ref{Intro}(a)]. The period is $a = 700$ nm, and the bar height and width are $h = 250$ nm and $w = 500$ nm, respectively. Similar GST metastructures fabricated by lithography were reported in the literature~\cite{Solomonov2023}.

The metasurface possesses the C$_2^v$ symmetry, which is known to support symmetry-protected BICs. %
Figure~\ref{Intro}(b) shows the TM mode dispersion and the magnetic field component $H_z$ distribution for both c- and a-GST. In a-GST, the field is strongly localized in the GST bars, while in c-GST, the field slightly leaks out of bars. The BIC resonant wavelength is about $1.55\,\mu\mathrm{m}$ (telecom C-band) in the amorphous state, and it red-shifts to $2.4\,\mu\mathrm{m}$ in the crystalline state, enabling reversible switching between the conventional 1550-nm telecom window and the emerging $2$-$\mu\mathrm{m}$ communications band for practical devices.

\begin{figure}[t]
\centering
\includegraphics[width = \linewidth]{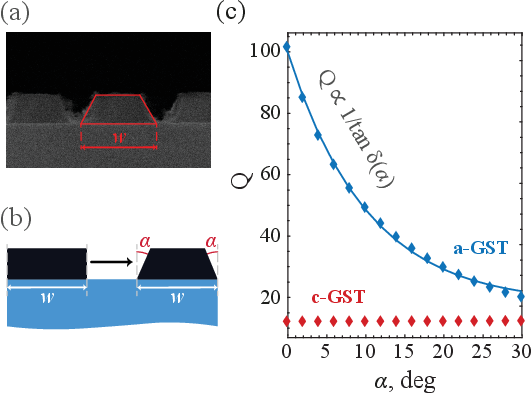}
\caption{(a) SEM image of a GST-based metasurface \cite{Solomonov2023}.
(b) Schematic illustration of the cross-section changing from rectangular to trapezoidal. Both inclination angles of the trapezoid (left and right) are equal to $\alpha$.
(c) Dependence of the $Q$ factor on $\alpha$ for crystalline and amorphous GST. The trend in the amorphous phase is largely governed by absorption.}
\label{Q(alpha)}
\end{figure}

During fabrication, the etching process takes a certain time, which leads to vertical non-uniformity of the structure. As a result, the GST bars deviate from the rectangular cross-section implied in the nominal design. Scanning electron microscopy (SEM) images reveal trapezoidal cross-sections of real structures, as illustrated in Fig.~\ref{Q(alpha)}(a). In addition, the cross-section fluctuates from bar to bar.

We start with a transition from a rectangle to a trapezoid preserving the C$_2^v$ symmetry of the metasurface.
We modify the bar cross-sections uniformly with a certain trapezoid inclination angle $\alpha$, as shown in Fig.~\ref{Q(alpha)}(b).
Then, we use numerical calculations to evaluate the influence of $\alpha$ on the $Q$ factor of q-BICs for c- and a-GST [Fig.~\ref{Q(alpha)}(c)]. 

{Moderate values of the $Q$ factor arise from the strong intrinsic losses of GST. At the wavelengths corresponding to the $\Gamma$ point, $\varepsilon_{a} = 16.8 + \mathrm{i} \cdot 0.24$ for amorphous phase and $\varepsilon_{c} = 43.8 + \mathrm{i} \cdot 3.73$ for crystalline phase. Despite the well-known advantages of high-$Q$ resonances, moderate-$Q$ q-BICs can be advantageous when the goal is to maximize coupling and field enhancement for robust operation, efficient excitation, and stable resonance characteristics for applications such as sensing, nonlinear photonics, switching, lasing, and others~\cite{Singh, Han2019, Liang2020, Fan:19, Aigner2022}. Therefore, the considered trapezoidal GST-based structure is well-suited for such applications, offering switching between the telecom C band and the emerging $2$-$\mu\mathrm{m}$ communication band through phase switching of GST.}

For c-GST, although the variation in $\alpha$ shifts the resonance wavelength of the q-BIC, the permittivity of c-GST is nearly constant in the range~\cite{Frantz2023Dec, Solomonov2023}.
Figure~\ref{Q(alpha)} reveals that the $Q$ factor remains essentially the same with negligible dependence on the angle $\alpha$.
This agrees with symmetry considerations, because an isosceles trapezoid does not reduce the C$_2^v$ symmetry, which is essential for the formation of symmetry-protected BICs.

In contrast, the permittivity dispersion of a-GST couples the resonance wavelength shift with a corresponding change in absorption loss \cite{Frantz2023Dec, Solomonov2023}.
{Therefore, we identify a mechanism for the $Q$ factor decrease that occurs despite the preservation of C$_2^v$ symmetry. In this mechanism, the $Q$ factor of the q-BIC decreases with changes in the nanostructure geometry (in this case, increasing $\alpha$) due to changes in the absorption loss accompyning the resonance shift induced by geometry perturbation.}

We quantify absorption using the loss tangent $\tan\delta = \mathrm{Im}(\varepsilon)/\mathrm{Re}(\varepsilon)$, where $\mathrm{Re}(\varepsilon)$ and $\mathrm{Im}(\varepsilon)$ are the real and imaginary parts of the permittivity. As we mentioned in the previous paragraph, the loss tangent becomes a function of the trapezoid inclination angle $\alpha$, $\tan{\delta} = \tan{\delta(\alpha)}$, due to absorption changes originating from the resonance frequency shift induced by geometry changes. 
Figure~\ref{Q(alpha)}(c) shows the approximation $Q \propto 1/\tan\delta$. This approximation agrees well with the numerical $Q$-factor dependence, showing the dominant impact of the absorption loss.

\section{Analytical model of fluctuating trapezoids}

\begin{figure}[th]
\centering
\includegraphics[width=\columnwidth]{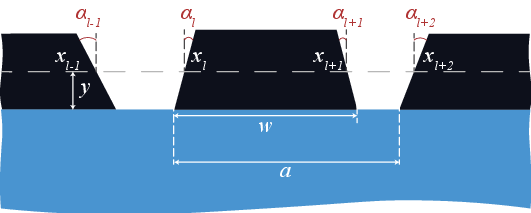}
\caption{Scheme of perturbed cross-sections of the structure with random disorder. Each bar has random sidewall inclination angles $\alpha_{\ell}$ and $\alpha_{\ell+1}$ uniformly distributed within a given range. Coordinates of the bar beginning and end at height $y$ are $x_\ell$ and $x_{\ell+1}$, respectively.}
\label{disscheme}
\end{figure}

In real structures, the inclination angle varies from bar to bar.
The disordered cross-sections are illustrated in Fig.~\ref{disscheme}.
Each bar has independently perturbed sidewall angles with random deviations
\begin{equation}
    \alpha_{\ell} = \alpha_0 \bigl(1 + R \cdot \sigma \bigr),
\end{equation}
where $R$ is a uniformly distributed random variable in the interval $\left[-1,\,1\right]$, $\sigma$ is the disorder amplitude, and $\alpha_0$ is the mean initial inclination angle.

We assume that structural disorder and material absorption contribute separately to the total $Q$ factor according to
\begin{equation}
\label{inverse law}
Q^{-1}_{\mathrm{tot}} = Q^{-1}_{\mathrm{dis}} + Q^{-1}_{\mathrm{abs}},
\end{equation}
where $Q_{\mathrm{tot}}$ is the total $Q$ factor, $Q_{\mathrm{dis}}$ is the disorder-induced $Q$ factor, and $Q_{\mathrm{abs}}$ is the absorption-induced $Q$ factor.
First, we analyze the structure with no material loss to study how angle fluctuations alone affect the $Q$ factor.

The structural fluctuations disrupt both the translation and C$_2^v$ symmetries, making Bloch theorem unsuitable for defining BIC conditions and transform the BIC into a q-BIC.
To address this issue, we employ a supercell approach \cite{Ni2017Mar} with the supercell period $Na$, where $N$ is the number of elements.

Angle fluctuations introduce a 2D disorder, i.e., the permittivity varies as a function of both $x$ and $y$.
Figure~\ref{disscheme} illustrates trapezoidal cross-sections with random angular fluctuations.
The base width $w$ and the distance $a$ between bar centers are fixed.
At a certain height $y$, the boundaries of the GST bar are given by the coordinates $x_\ell(y)$ and $x_{\ell+1}(y)$, with $\ell$ taking even values $0,\ldots, 2N-2$.
Since permittivity remains periodic in $x$ with a period $Na$, it can be expanded into Fourier series as
\begin{equation}
\label{eq:fractial_epsilon}
\varepsilon(x, y) = \sum_{m \in \mathbb{Z}} \varepsilon_m(y) e^{i m \frac{2 \pi}{Na} x},
\end{equation}
where $\varepsilon_m(y)$ are the Fourier coefficients
\begin{multline}
\label{eq:fractial_Bloch}
\varepsilon_{m}(y) =
\frac{1}{Na}\left[
\sum_{\ell} \int_{x_\ell}^{x_{\ell+1}} \varepsilon\, e^{-i m \frac{2 \pi}{Na}x} \, dx
+ \right.\\
\left. \sum_{\ell} \int_{x_{\ell+1}}^{x_{\ell+2}} e^{-i m \frac{2 \pi}{Na}x} \, dx
\right],
\end{multline}
where $\varepsilon$ is the GST permittivity, and the permittivity of air is set to 1.

To evaluate the Fourier coefficients, we note that the integration interval is partitioned into subintervals with constant permittivity, where the subinterval boundaries are random.
We use the following parametrization
\begin{equation}
\begin{cases}
x_\ell = a \cdot \ell/2 + y \cdot \tan{\alpha_{\ell}}, \\
x_{\ell+1} = a \cdot \ell/2 + w - y \cdot \tan{\alpha_{\ell+1}}
\end{cases}
\end{equation}
to evaluate those integrals.

Assuming a small-angle approximation ($\tan{\alpha_l} \approx \alpha_l \ll 1$), we obtain (see Appendix ~\ref{Fourier_coefficients_of_dielectric_permittivity} for detailed derivation)
\begin{multline}
\label{eq:epsilon_expansion_1}
\varepsilon(x, y) = \frac{(\varepsilon - 1)w + a}{Na}  \sum_{m}  \sum_{\ell} e^{i \frac{2\pi m}{Na} (x - a\ell)}
\\
+ \underbrace{\frac{(1 - \varepsilon)}{N} \sum_{m}  \sum_{\ell} \Delta \rho_{\ell} (y) e^{i \frac{2\pi m}{Na} (x - a\ell)}}_{\Delta \varepsilon(\mathbf{r})},
\end{multline}
where we introduce the notation $\Delta\rho_{\ell}(y) = y(\alpha_{\ell} + \alpha_{\ell+1})/a$. The first term corresponds to the unperturbed permittivity, and the second term defines the perturbation $\Delta \varepsilon(\mathbf{r})$.

The radiative contribution to the $Q$ factor due to coupling between a BIC and leaky resonant modes is evaluated as \cite{Koshelev2018Nov}
\begin{equation}
Q_{\mathrm{dis}}^{-1} = \frac{\omega_{\mathrm{n_0}}^{(0)}}{2}\,\mathrm{Im}\left(\sum_{m} \frac{V_{\mathrm{n_0 m}}^2}{\omega_{\mathrm{m}}^{(0)} - \omega_{\mathrm{n_0}}^{(0)}}\right),
\end{equation}
where $\omega_{\mathrm{n_0}}^{(0)}$ is the unperturbed (purely real) BIC frequency of the state $\bf{E}_{\mathrm{n_0}}^{(0)}$, $\omega_{\mathrm{m}}^{(0)}$ are the complex frequencies of the continuum basis states $\bf{E}_{\mathrm{m}}^{(0)}$, and
\begin{equation}
V_{\mathrm{n_0 m}} = \int \mathrm{d}\mathbf{r}\, \Delta \varepsilon(\mathbf{r})\, \bf{E}_{\mathrm{n_0}}^{(0)} \bf{E}_{\mathrm{m}}^{(0)}.
\end{equation}
The $Q_{\mathrm{dis}}$ behavior is governed by the mean value $E\!\left[\Delta \varepsilon^2 (\mathbf{r})\right]$, where the entire dependence on the trapezoid angles (over which we average) is contained in $\Delta \rho_{\ell} (y)$.

Then, we use the expression for the mean values weighted by the probability density. Here, assuming independent uniform distributions,
$\alpha_{\ell} \in [\alpha_0 - \delta\alpha_0, \alpha_0 + \delta\alpha_0], \delta\alpha_0 = \sigma\alpha_0$,
the probability density is
\begin{equation}
p =\frac{1}{2\delta\alpha_0}=\frac{1}{2\sigma\alpha_0}.
\end{equation}
Meanwhile, in this derivation, we need to evaluate the mean values of types: $E[\alpha_{\ell}^2]$ and $E[\alpha_{\ell} \alpha_{\ell^{'}}] = E[\alpha_{\ell} ] E[\alpha_{\ell^{'}}]$. Their values are
\begin{equation}
E[\alpha_{\ell} \alpha_{\ell^{'}}]=\alpha_0^2, \qquad
E[\alpha_{\ell}^2]=\alpha_0^2 + \frac{(\delta\alpha_0)^2}{3}
= \alpha_0^2\left(1+\frac{\sigma^2}{3}\right).
\end{equation}
This yields the inverse quadratic scaling
\begin{equation}
Q_{\mathrm{dis}} \propto \sigma^{-2},
\end{equation}
and, at fixed $\sigma$,
\begin{equation}
Q_{\mathrm{dis}} \propto \alpha_0^{-2}.
\end{equation}

\section{Numerical results}

Now we examine $Q_{\mathrm{dis}}$ numerically. We use a finite element method to calculate the total $Q$ factors of the disordered structure with a supercell comprising $N = 20$ periods.
Data are collected from an ensemble of $M=100$ structures for each value of the disorder amplitude $\sigma$.
The computed $Q$ factor is averaged over the ensemble as
\begin{equation}
\langle Q_{\mathrm{tot}} \rangle = \frac{1}{M} \sum_{i=1}^{M} Q_{\mathrm{tot}}^{(i)},
\end{equation}
The inclination angle fluctuates around $\alpha_0 = 5^{\circ}$ and is uniformly distributed within the interval $[\alpha_0(1-\sigma),\,\alpha_0(1+\sigma)]$.
We vary the disorder amplitude $\sigma$ from 0.01 to 0.1 in steps of 0.01.

The focus is on the dependence of the $Q$ factor on $\sigma$, since the fluctuation amplitude is typically used to assess the $Q$ factor behavior in experiments \cite{Barreda2022Aug}.
The dependence of the $Q$ factor on $\alpha_0$ at fixed $\sigma$ is discussed in Appendix~\ref{Q_factor_dependencies_on_forming_angle_with_fluctuations}.

\begin{figure}[htbp]
\centering
\includegraphics[width=\columnwidth]{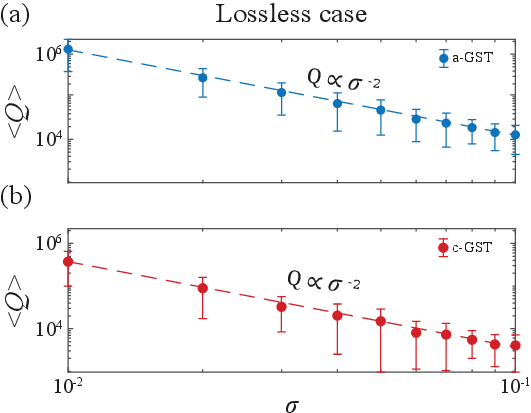}
\caption{
$Q$ factors as functions of the disorder amplitude $\sigma$ in (a) crystalline and (b) amorphous GST without material loss. Bars indicate the spread of $Q$ factors for each $\sigma$. Lines correspond to inverse quadratic fits.}
\label{Q(sigma)_cryst_no_losses}
\end{figure}

Initially, we consider the structure with no material loss. We approximate the dependence of the mean $Q$ factor on parameter $\sigma$ with quadratic relationship $\langle Q \rangle \propto \sigma^{-2}$. Figure~\ref{Q(sigma)_cryst_no_losses} shows that for both c-GST and a-GST, the dependence of the $Q$ factor on $\sigma$ agrees well with an inverse quadratic scaling, which is consistent with the analytical prediction.
In particular, the $Q$ factor is larger for the amorphous phase than for the crystalline one, which can be attributed to the reduced permittivity contrast between a-GST and the SiO$_2$ substrate~\cite{Vincenti2024Aug}.

\begin{figure}[h!]
\centering
\includegraphics[width=\columnwidth]{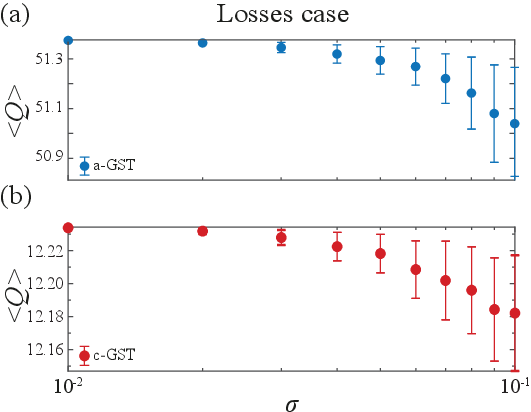}
\caption{Dependence of the $Q$ factor on the disorder amplitude $\sigma$ in (a) crystalline and (b) amorphous GST with material loss. Bars indicate the spread of $Q$ factors for each $\sigma$. The $Q$ factor weakly depends on $\sigma$.}
\label{Q(sigma)_cryst_losses}
\end{figure}

Next, we consider the impact of material loss.
As shown in Fig.~\ref{Q(sigma)_cryst_losses}, the $Q$ factors barely change with $\sigma$ for both crystalline and amorphous phases.
These $Q$ factors are comparable to those observed for the identical trapezoids with $\alpha = 5^{\circ}$ [Fig.~\ref{Q(alpha)} (c)] and differ by several orders of magnitude from the $Q$ factors for random shape fluctuations with no material loss.
This discrepancy arises because the total $Q$ factor is affected by both $Q_{\mathrm{abs}}$ and $Q_{\mathrm{dis}}$, which contribute according to the inverse law (see Eq.~\eqref{inverse law}).
In the considered regime, $1/Q_{\mathrm{abs}}$ dominates, and the contribution from $Q_{\mathrm{dis}}$ is negligible
\begin{equation}
Q^{-1}_{\mathrm{tot}} = Q^{-1}_{\mathrm{dis}} + Q^{-1}_{\mathrm{abs}} \approx Q^{-1}_{\mathrm{abs}}.
\end{equation}

Since fluctuations are small, the resonance frequency changes weakly, and absorption remains nearly constant resulting in a weak dependence of the $Q$ factor on $\sigma$. {These findings establish that in both GST phases, BICs are robust to disorder in trapezoidal cross-sections.}

\section{Conclusion}

We have quantified the impact of lithography-induced geometric imperfections on the $Q$ factor of symmetry-protected BICs in a GST-based metasurface. We have uncovered the interplay between material loss, dispersion, and lithographic disorder, elucidating their combined effect on the $Q$ factor and potential applications of this GST-based metasurface.

We have considered trapezoidal deviations from the nominal rectangular cross-section of bars in both the amorphous and crystalline phases of GST, employing a comprehensive approach that combines analytical derivations with finite-element simulations.
For identical isosceles trapezoids preserving the $C_2^v$ symmetry of the unit cell, the $Q$ factor in the crystalline phase remains nearly constant, confirming the topological robustness of symmetry-protected BICs against uniform shape deformations. 
In contrast, for the amorphous phase, we have identified a mechanism where the the $Q$ factor degrades with the trapezoidal angle. This effect originates from absorption changes that accompany the resonance shift, following an inverse scaling law $Q \propto(\tan\delta)^{-1}$.
{This mechanism can also arise in other geometries with a significant dispersion near the BIC wavelength.}
For random element-to-element variations of the trapezoid angle, the $C_2^v$ symmetry is broken, and the BICs are transformed into finite-$Q$ resonances. Without material loss, the
$Q$ factor follows an inverse-quadratic scaling with disorder, $Q_\text{dis} \propto \sigma^{-2}$ and $Q_\text{dis} \propto \alpha_0^{-2}$, which has been obtained analytically and then verified numerically. {More generally, the obtained inverse-quadratic relationship holds regardless of the material where trapezoidal disorder is induced. It remains valid for any high-index dielectric metasurface platform where the main manufacturing defect is variations in the sidewall angle of a unit element.} 
When intrinsic material losses are essential, absorption dominates the total $Q$ factor, and its dependence on geometric disorder becomes negligible. This renders BICs supported by this GST-based metasurface robust against trapezoidal fluctuations.

 {From a technological perspective, these findings indicate that GST-based BIC metasurfaces are capable of operating within realistic fabrication tolerances across both the telecom C-band (amorphous phase $\sim$1550~nm) and the emerging 2-$\mu$m communications window (crystalline phase). In particular, for lossy phase-change materials, the achievable $Q$ factor is ultimately limited by material absorption rather than by geometric disorder. Thus, the demonstrated disorder robustness of the considered BICs, combined with the phase-switching capability of GST, opens a route toward practical reconfigurable photonic devices such as tunable filters, optical switches, modulators, and sensing platforms.}

\section*{Acknowledgments}
This work was supported by the Russian Science Foundation (Grant No.~24-72-10038). The authors thank Lydia Pogorelskaya for proofreading the manuscript.

\appendix
\onecolumngrid

\section{Fourier expansion}
\label{Fourier_coefficients_of_dielectric_permittivity}

The Fourier coefficients of the array permittivity are calculated as
\begin{equation}
\label{eq:fractial_Bloch2}
\varepsilon_{m}(y) =
\frac{1}{Na}\left[
\sum_{\ell=0}^{2N-2} \int_{x_\ell}^{x_{\ell+1}} \varepsilon\, e^{-i m \frac{2 \pi}{Na}x} \, dx
+
\sum_{\ell=0}^{2N-2} \int_{x_{\ell+1}}^{x_{\ell+2}} e^{-i m \frac{2 \pi}{Na}x} \, dx
\right],
\end{equation}
where $\varepsilon$ is the permittivity of GST, and the permittivity of air is set to 1.

Let us first calculate the integrals for constant permittivity of bars, assuming small-angle approximation ($\tan{\alpha_l} \approx \alpha_l \ll 1$)
\begin{multline}
\int_{x_\ell}^{x_{\ell+1}} \varepsilon e^{-i m\frac{2 \pi}{Na} x} \, dx
=
-\frac{\varepsilon Na}{i 2\pi m} e^{-i \frac{\pi m}{N} \ell}
\left(
e^{-i \frac{2 \pi m}{Na} (x_\ell + w -y(\alpha_{\ell} + \alpha_{\ell+1}))}
-
e^{-i \frac{2 \pi m}{Na} y \alpha_{\ell}}
\right) \\
=
\frac{i \varepsilon Na}{2\pi m} e^{-i \frac{\pi m}{N} \ell}
\left(
e^{-i \frac{2 \pi m}{Na} (w - y \alpha_{\ell+1})}
-
e^{-i \frac{2 \pi m}{Na} y \alpha_{\ell}}
\right).
\end{multline}

Now, we expand the exponents into a Taylor series:
$e^{-i\frac{2 \pi m}{Na} y \tan{\alpha_{\ell}}}$ and $e^{-i\frac{2\pi m}{Na}(w - y \alpha_{\ell+1})}$.
Therefore,
\begin{multline}
\int_{x_\ell}^{x_{\ell+1}} \varepsilon e^{-i m\frac{2 \pi}{Na} x} \, dx
\approx
\frac{i \varepsilon Na}{2\pi m} e^{-i \frac{\pi m}{N} \ell}
\left[
1 - i \frac{2 \pi m}{Na} w + i \frac{2 \pi m}{Na} y \alpha_{\ell+1}
- 1 + i \frac{2 \pi m}{Na} y \alpha_{\ell}
\right] \\
= \varepsilon w e^{-i \frac{\pi m}{N} \ell} - \varepsilon y (\alpha_{\ell+1} + \alpha_{\ell}) e^{-i \frac{\pi m}{N} \ell}.
\end{multline}

Similarly, for the air region
\begin{equation}
\int_{x_{\ell+1}}^{x_{\ell+2}}  e^{-i m\frac{2 \pi}{Na} x} \, dx
\approx
- w e^{-i \frac{\pi m}{N} \ell} + y e^{-i \frac{\pi m}{N} \ell} (\alpha_{\ell+2} + \alpha_{\ell+1}).
\end{equation}

Thus,
\begin{equation}
\varepsilon_{m}(y) = \frac{(\varepsilon - 1)w + a}{Na} \sum_{\ell=0}^{2N-2} e^{-i \frac{\pi m}{N} \ell}
+
\frac{(1 - \varepsilon)y}{Na}  \sum_{\ell=0}^{2N-2} (\alpha_{\ell} + \alpha_{\ell+1}) e^{-i \frac{\pi m}{N} \ell}.
\end{equation}

Introducing $\Delta\rho_{\ell}(y) = y(\alpha_{\ell} +\alpha_{\ell+1})/a$, we obtain
\begin{equation}
\label{eq:epsilon_expansion}
\varepsilon(x, y) = \frac{(\varepsilon - 1)w + a}{Na}  \sum_{m}  \sum_{\ell=0}^{2N-2} e^{i \frac{2\pi m}{Na} (x - a\ell)}
+
\frac{(1 - \varepsilon)}{N} \sum_{m}  \sum_{\ell=0}^{2N-2} \Delta \rho_{\ell} (y) e^{i \frac{2\pi m}{Na} (x - a\ell)}.
\end{equation}

\section{$Q$ factor dependencies on inclination angle with fluctuations}
\label{Q_factor_dependencies_on_forming_angle_with_fluctuations}

\begin{figure}[t]
\centering
\includegraphics[width = 0.5\linewidth]{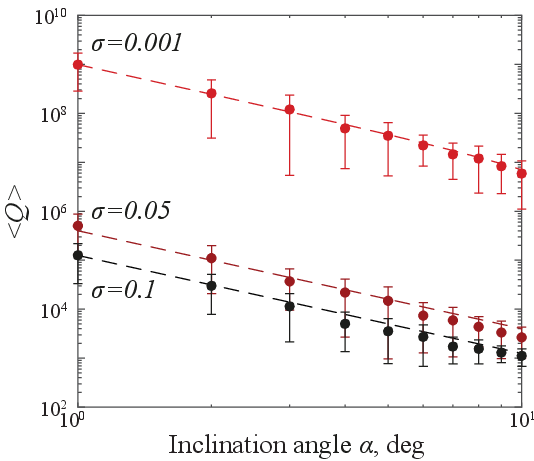}
\caption{$Q$ factor dependencies (log--log scale) on the mean inclination angle $\alpha_0$ in crystalline GST without material loss. Three series correspond to $\sigma=0.001,\,0.05,\,0.1$ from top to bottom. Lines correspond to inverse-quadratic fits.}
\label{difsigma}
\end{figure}

For three fixed disorder amplitudes ($\sigma=0.001,\,0.05,\,0.1$), we calculated the dependencies of the $Q$ factor on the mean inclination angle $\alpha_0$ in the presence of fluctuations.
We consider only the crystalline phase of GST without material loss.
Figure~\ref{difsigma} shows that for all fixed $\sigma$, the inverse-quadratic scaling of the $Q$ factors is preserved.

\end{document}